# Electron-lattice coupling and partial nesting as the origin of Fermi-Arcs in manganites

Juan Salafranca,[1,2] Gonzalo Alvarez,[3] and Elbio Dagotto[1,2]

[1]*Department of Physics and Astronomy, University of Tennessee, Knoxville, Tennessee 37996, USA*
[2]*Materials Science and Technology Division, Oak Ridge National Laboratory, Oak Ridge, Tennessee 32831, USA*
[3]*Computer Science and Mathematics Division and Center for Nanophase Materials Science,
Oak Ridge National Laboratory, Oak Ridge, Tennessee 37831, USA*
(Dated: May 26, 2010)

A tight-binding model for $e_g$ electrons coupled to Jahn-Teller lattice distortions is studied via unbiased Monte-Carlo simulations. By focusing on the periodicity of the Jahn Teller distortions, and the one-particle spectral function, our results clarify the physical origin of the Fermi-arcs phase observed in layered manganites. In a range of parameters where no broken symmetry phase exists, the nearly nested Fermi surface favors certain correlations between Jahn Teller distortions. The spectral weight near the Brillouin zone edge is suppressed, leading to the pseudogap in the density of states. We discuss the stability of this phase as a function of temperature and coupling strength for different hole dopings.

## I. INTRODUCTION

The pseudogap phase, where the Fermi surface appears as four disconnected arcs, has been considered as a crucial ingredient to understand the physics of high critical temperature superconductors based on copper oxide layers. Recently, angle resolved photoemission spectroscopy (ARPES) experiments[1] in a layered manganite revealed a strikingly similar Fermi surface, unveiling unexpected similarities between two of the most studied strongly correlated families of materials, the manganites and cuprates. The results in Ref. 1 are an example of the renewed interest in layered manganites,[2–5] encouraged by these similarities, and made possible by the continuously improving crystal quality, and the advances in experimental techniques.

One of the attractive features of manganites is the possibility to dope holes in the system by chemical substitution of, for instance, trivalent La by divalent Sr.[6] Another parameter that can vary is the dimensionality: $MnO_6$ octahedra form a three dimensional (3D) network in manganites with the perovskite structure (as $LaMnO_3$), but they are also found forming a two dimensional (2D) layer, one octahedra thick, separated by rock-salt type layers in single layer manganites ($LaSrMnO_4$), and as two octahedra thick layers in bi-layer manganites ($La_2SrMnO_7$). Sample quality has allowed a recent careful and precise determination of the phase diagram of the bilayer compounds as a function of doping[7]. It is interesting to remark that physical properties are so sensitive to doping, that these properties, together with theoretical considerations are sometimes used to improve precision in the determination of the number of carriers.[8]

The bilayer with hole doping $x$=0.4 is an interesting compound, and its properties have been very well established.[9,10] Similarly to some of the perovskite 3D compounds, it displays a temperature induced ferromagnetic metal to paramagnetic insulator transition, and the concomitant colossal magnetoresistance effect under applied field. It is for this compound that the mentioned ARPES experiments were carried out.[1] They showed a peculiar Fermi surface reminiscent of the pseudogap cuprates, with suppressed spectral weight in the $(\pi,\pi) \rightarrow (0,\pi)$ antinodal direction, coexisting with a coherent quasiparticle peak in the $(\pi,\pi) \rightarrow (0,0)$ nodal direction. A strong reduction *of at least 90%*[3] (as opposed to the complete suppression[1]) had also been reported. The very fact that such a controversy can be held shows the precision reached by photoemission techniques.

The phase diagram as a function of doping, includes different magnetic ground states,[7] but as metallicity survives for a wide doping range, other Fermi surface measurements with ARPES have been done for different values of $x$. Subtle and precision demanding effects, such as bilayer splitting of the Fermi surface (due to hopping of the carriers between the two Mn planes in the bilayers) have been observed at hole dopings lower than $x$=0.4.[11,12] Experiments for higher dopings can be found in Ref. 5. In general, with today's ARPES precision it is possible to track changes and detailed features of the spectral function and the Fermi surface, making this technique specially suitable to study the different states in the metal to insulator transitions, or the transition itself.[4] Recent efforts complement the pioneer one[13] where the reduction in the density of states at the Fermi energy, then termed pseudogap, was first observed in the context of manganites.

This numerical study builds over a long experience on double exchange model coupled to phonons,[14–17] with the aim of establishing direct connection between results of the microscopic model and experiments. In order to achieve this goal we have used twisted boundary conditions for the fermionic sector (Section II A), a well understood tool but not priorly used (to our knowledge) in numerical works similar to the present one. This technical improvement provides a much more detailed information about the spectral function, both at the Fermi level and other energies. The results thus obtained, show that a microscopic model with large Hund's coupling and $e_g$ electrons coupled to cooperative Jahn-Teller distortions is enough to understand the pseudogap phase in manganites, and that this phase, indeed displays a cuprate-like Fermi surface in agreement with experiments.[1,3] Fluctuations in the Jahn-Teller distortions at different sites are correlated, the wave vectors are such that they connect nested (as proposed in Ref. 3) or nearly-nested parts of the Fermi surface. This produces a loss of coherence and spectral weight for electrons with the nested momenta, near the Brillouin zone face, giving rise to the arc-like Fermi surface.

The rest of the paper is organized as follows. Section II A explains the microscopic model, the different approximations, the Monte Carlo procedure, and the use of twisted boundary conditions. The first results, regarding the model without interactions are found in Section III, where the concept of non-perfect (or partial) nesting is defined. Section IV contains the results of numerical simulations for the half doped case, in particular, the effects of electron-phonon coupling and temperature in the spectral function, and the changes they induce in the Fermi surface. Section V extends the analysis to other dopings, $x=0.4$ and $x=0.6$. The conclusions and a brief summary are presented in Section VI. The Appendix explains some details about the electron-phonon coupling in manganites.

## II. MODEL

### A. Layered Manganites

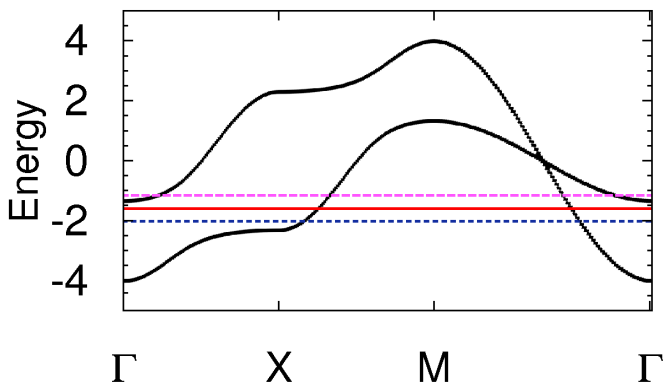

FIG. 1: Energy Bands for the 2D, two orbital model in the ferromagnetic state, with no electron-phonon coupling (Eq. (2)). The different horizontal lines correspond to the chemical potentials for $x=0.4$ (pink, slashed), $x=0.5$ (red, continuous), and $x=0.6$ (blue, dashed), the three different dopings studied in this work.

The anisotropy in the conductivity of the bilayer manganites is large ($\gtrsim 10^2$),[9] and although some weak interlayer coupling exists, assuming a two dimensional system is a good starting point. The focus of this article is not on determining ordering temperatures, but on the electronic structure of the system at low temperatures. Also, only one MnO$_2$ plane is considered. There are differences between single layer and bi-layer manganites, but this work shows that this approximation gives a good understanding on the physics of bi-layered manganites.

The active degrees of freedom are the Mn $d$ orbitals. Among them, the $t_{2g}$ orbitals lie deeper in energy due to crystal field splitting, and the Fermi surface arises from the $e_g$ orbitals, $x^2-y^2$, and $3z^2-r^2$. A commonly made approximation is to take Hund's coupling to infinity, and to consider the 3/2 spin resulting from the $t_{2g}$ electrons as a classical spin. In general, large Hund's coupling leads to the double exchange model, and to very interesting physics that has been much explored in the past. Our focus will be in temperatures much lower that the Curie temperature of the system ($T_c = 120$ K for $x=0.4$) and, as a consequence, we will assume perfect ferromagnetic order within the MnO$_2$ plane. Therefore, here perfectly spin polarized electrons move in a ferromagnetic background. The $t_{2g}$ background spins only add a constant term to the Hamiltonian, and it is possible to ignore both them and the spin degree of freedom of the perfectly polarized $e_g$ electrons. Previous extensive numerical work by several groups, and direct comparison between our results and ARPES experiments, justify the approximations made.

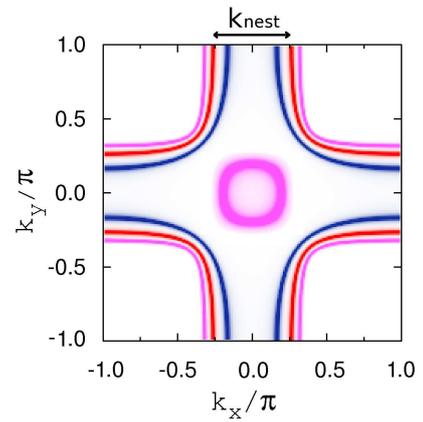

FIG. 2: Fermi surface for the same dopings as in Fig. 1 $x=0.6$ (blue, dark gray), $x=0.5$ (red, medium gray), and $x=0.4$ (pink, light gray). The *better nesting* (i.e. smaller angle between tangents of different pieces of the Fermi surface) for $x=0.4$, as compared to $x=0.6$, implies that a smaller electron-phonon coupling is needed to induce the pseudogap phase. For $x=0.5$ the nested portions of the Fermi surface are separated by $k_{nest} \approx 2\pi/4a$, which is commensurate with the systems size that can be studied with the Monte Carlo technique.

We are aware that charge ordered (CO) states are the ground states at $x=0.5$,[8] and $x=0.6$,[7] dopings also studied in this paper. But these states are fragile and as $x$ changes as little as 2% they become A-type antiferromagnetically ordered and display metallic conductivity.[7] In the A-type antiferromagnetic state the spins are FM aligned within the planes, with antiferromagnetic coupling in the $z$ direction. Note that in this situation, the double exchange[18] mechanism allows the electrons to move only in the $xy$ plane. The results presented in this paper for $x=0.5$ and $x=0.6$ assuming FM order are not only useful to gain deeper understanding, but they are also expected to be relevant for antiferromagnetic metallic samples with hole doping close to (but not exactly equal to) these values.

### B. Hamiltonian

The Hamiltonian, therefore, describes the $e_g$ electrons, their kinetic energy in a ferromagnetic background, and the interaction with the lattice distortions:

$$H = H_K + H_{e-ph} \tag{1}$$

The kinetic energy term is a tight-binding Hamiltonian:

$$H_K = H = - \sum_{i,u,a,a'} t^u_{a,a'} C^+_{i,a} C_{i+u,a'}, \qquad (2)$$

where $a$ and $a'$ run over orbitals $|1\rangle = |x^2-y^2\rangle$, and $|2\rangle = |3z^2-r^2\rangle$, $t$ takes into account the different overlaps between the orbitals along the directions u=x,y:[19] $t^x_{1,1}=3t^x_{2,2}=\sqrt{3}t^x_{1,2}=t$ and $t^y_{1,1}=3t^y_{2,2}=-\sqrt{3}t^y_{1,2}=t$. $t$ is the energy unit throughout this work. The close similarity between the spectral function arising from our calculation and ARPES experiments allow us to estimate $t$ to be 0.5-0.6 eV. The units are such that the Boltzmann constant $K_B$ is one. Together with the estimated value of t, T=0.01 in this work is approximately 60K.

Freezing the magnetic degree of freedom allows us to concentrate in the role of the lattice. The second part of the Hamiltonian describes lattice distortions and their interaction with electrons. Electrons are adiabatically coupled to those distortions, and the kinetic aspects or the quantum nature of phonons are not taken into account. Furthermore, we will concentrate on the Jahn-Teller distortions,[20] ignoring the small tetrahedral distortion of the lattice and buckling modes that change the Mn-O-Mn angles. The phononic part of the Hamiltonian then reads:

$$H_{e-ph} = \lambda \sum_i \left( -Q_{1i}\rho_i + Q_{2i}\tau_{xi} + Q_{3i}\tau_{zi} \right)$$
$$+ \frac{1}{2} \sum_i \left( Q^2_{1i} + Q^2_{2i} + Q^2_{3i} \right) , \qquad (3)$$

where $\rho_i$ is the density operator at site $i$, and $\tau_{xi}$ and $\tau_{zi}$ are the corresponding Pauli matrices in the $e_g$ subspace, that express the coupling of distortions, $Q$, to electrons. In particular, $Q_{1i}$ is the breathing mode of the $MnO_6$ octahedra around the $i$-th manganese ion, $Q_{2i}$ and $Q_{3i}$ are the Jahn Teller modes. Cooperative effects are taken into account by expressing the $Q$'s in terms of the positions on the oxygen ions, as explained in the Appendix. The first part of Eq. (3) corresponds to the electron-phonon interaction, while the second part is the elastic energy cost of the distortions. $\lambda$ measures the interaction strength. The Hamiltonian (3) is well known.[21] In several prior efforts, the sign of the coupling between the breathing mode and the density was positive. We briefly argue why it should be negative (as in Eq. (3)), and the (limited) effect that this has on the physics of manganites, in the Appendix, where further details about the model can be found. This difference in sign does not affect the conclusions of previous efforts that included the breathing mode since usually these modes were suppressed by a large spring constant. Electron-electron interaction is also important in manganites, especially at low hole doping. Although interorbital and intraorbital electron-electron interactions exist, the large Hund's coupling approximation made here (and in other studies of manganites[21]) only takes into account the intraorbital part of the interaction. We expect interorbital interactions to favor orbital order. These effects become smaller as hole doping increases, but anyhow the $\lambda$ in Eq. (1) should be understood as an effective parameter.

### C. Twisted boundary conditions and computational details.

In this work, the focus is on the electronic structure and the periodicity and magnitude of the lattice distortions. These properties are studied by means of Monte Carlo simulations with twisted boundary conditions.

The use of twisted boundary conditions allows us to obtain a good description of the electronic spectral function. The approximation of considering lattice distortions with a periodic system of size L is still made. But for a correlated system (and any electronic system in general) which can be mapped onto a one-electron system coupled to a periodic classical field, as in our case, the use of twisted boundary conditions constitutes a systematic way to improve the accuracy of the calculations.

Since within the simulation system, distortions have period L, the potential *felt* by electrons has not completely lost the translation symmetry, and Bloch theorem can be applied. It is then possible to choose the eigenstates, $\psi_{n\mathbf{k}}$ so that:

$$\psi_{n\mathbf{k}}(\mathbf{r} + \mathbf{R}) = \exp i(\mathbf{k}.\mathbf{R}) \psi_{n\mathbf{k}}(\mathbf{r}) , \qquad (4)$$

where $\mathbf{R}$ are the vectors of the Bravais lattice with the periodicity of the system, in 2D, $\mathbf{R}=(n_x L, n_y L)$. If we consider the subspace of wave functions that obey $\phi(\mathbf{r}+\mathbf{R})=\exp i(\mathbf{k}.\mathbf{R})\phi(\mathbf{r})$ for a particular $\mathbf{k}'$, the states $\psi_{n\mathbf{k}'}$ are then a basis for that subspace. But since $\psi_{n\mathbf{k}'}$ are also eigenstates, it follows that the Hamiltonian operator does not mix functions belonging to different subspaces. The Hamiltonian is box diagonal in any basis with well defined periodicity. This is computationally very convenient, as we can diagonalize each subspace and calculate any operator that is a function only of the Hamiltonian (for example, the Green and spectral functions) within that subspace.

In particular, for a periodic tight binding Hamiltonian in a real space basis, like Eq. (2), all the blocks in the diagonal $\hat{H}(\mathbf{k})$ are equal, except for the elements connecting sites across the boundaries. For these we have:

$$\langle \mathbf{r}_i | H(\mathbf{k}) | \mathbf{r}_j \rangle = \exp i\mathbf{k}.\mathbf{R} \langle \mathbf{r}_i | H | \mathbf{r}_j - \mathbf{R} \rangle. \qquad (5)$$

A systematic way to chose $\mathbf{k}$ is $k_\alpha = 2\alpha\pi/M$, with $\alpha=0,1,...,M-1$, and this is the choice made in the present work. Therefore, a 12×12 lattice with 8×8 twisted boundary conditions (L=12 and M=8), implies considering distortions within a 12×12 lattice, calculating the Hamiltonian (1), and solving the 64 different blocks with boundary conditions for $x$ and $y$, as in Eq. (5). Notice that from the point of view of the electronic structure, this is completely equivalent to solving the Hamiltonian with the same classical field of period L in an LxM lattice. That is easy to understand just by considering any complete set of wave functions (being it the basis in real space or the eigenvectors) in the LxL lattice, and extending them to the (LxM)x(LxM) lattice in the obvious way: $\psi_{nk}(r_l) \to \psi_{nk}(r) = \exp r_m \frac{2\pi i}{M} \psi_{nk}(r_l)$, with $r = r_l + L r_m$.

In our Monte Carlo simulations, the acceptance of rejection of configurations has been calculated with periodic boundary conditions, while the determination of the thermal-averaged physical quantities has been obtained using twisted boundary conditions. Systems sizes are 12×12 Mn sites, and 8×8 phases are used for the twisted boundary conditions. Typically, the thermalization process needed 2000 Monte Carlo steps, and the Markov chains were 7000-15000 steps long (measurements were taken every 10 steps). The usual definition of Monte Carlo step and further details about the Monte Carlo procedure can be found in Ref. 22.

Finite size effects are checked with Monte Carlo simulations on 8×8 systems and, for $T$=0, the direct optimization of the oxygen positions in larger systems (up to 30×30). These size effects are small, as discussed in Section IV B.

Along the manuscript, we present results about the spectral function A($\mathbf{k}$,$\omega$) and the thermal average of the square of the distortions in $\mathbf{k}$-space, $\langle Q^2(\mathbf{k})\rangle$. The unit in $\mathbf{k}$-space is one over the lattice spacing, so that the Brillouin zone goes from -$\pi$ to $\pi$, both in $x$ and $y$. For the distortions, we calculate the correlation function of the $\alpha$=1,2,3 $Q_\alpha$ modes explained in the Appendix: $Q_\alpha(\mathbf{r}) Q_\alpha(\mathbf{r},\mathbf{r}+\delta)$. After averaging over the lattice sites, $\mathbf{r}$, we Fourier transform the correlation function. A straight forward calculation shows that the result is a thermal averaged structure factor for each of the distortion modes,

$$Q_2^2(\mathbf{k}) = \sum_n exp\left(E_n/K_BT\right) Q_{2,n}^2(\mathbf{k}) , \quad (6)$$

and similarly for other modes. We have dropped the brackets of thermal average in the notation for convenience.

## III. UNDISTORTED SYSTEM AND NON-PERFECT NESTING

The band diagram for the system with no electron-phonon coupling is shown in Fig. 1. Within the non-interacting system, changing doping only changes the chemical potential. Figure 2 shows the Fermi surfaces for different dopings. Already in this limit there are some interesting features. It is remarkable how it reproduces the essential topology of the Fermi surface of layered manganites, as compared to experimental results,[1,11] and LDA calculations.[2] For $x$=0.4 there is a small electron pocket, also found in experiments at this doping.[1,3] We can ignore it for now, and concentrate on the features that are common to all dopings. The Fermi surface for these two dimensional systems at the dopings considered is a closed curve *centered* at the M point, which appears as four pieces in Fig. 2. Of course, the Fermi surface presents the four-fold symmetry imposed by the lattice, since there are no interactions to remove it. Symmetry allows $k$ along the ($\pi$,$\pi$)→(0,0) and ($\pi$,$\pi$)→(0,$\pi$) directions to be different, and so they are. For all dopings in Fig. 2 the curvature along the diagonal is clearly larger (small curvature radius) than the curvature along the ($\pi$,$\pi$)→(0,$\pi$) direction. These essential features of the Fermi surface are not only characteristic of layered manganites, but they are also shared by the high-$T_c$ cuprates, which also share the effective two dimensional electron system and the square lattice. Both for convenience and to remark the similarity of the Fermi surfaces of the two families of compounds, we will follow Ref. 1, and refer to the two directions as nodal ($\pi$,$\pi$)→(0,0) and antinodal ($\pi$,$\pi$)→(0,$\pi$) directions.

Nesting takes place when finite parts of the Fermi surface are connected by one wave vector. It is well known that in these situations the system might be unstable with respect to a perturbation with the periodicity of that wave vector, even when the coupling between the electrons and the perturbation is small. Different and numerous examples exist and are well understood, such as structural (Peierls) instabilities and Spin and charge density waves (see, for instance, Ref. 23 and references therein). However, this is not the situation here. As shown if Fig. 2, the curve, when drawn around the M point, has constant-sign non-zero curvature and, therefore, a particular vector in $k$ space can only connect two points (times fourfold degeneracy) of the Fermi surface: there is no perfect nesting in layered manganites for the dopings we are considering.

However, although non-zero, the curvature of the Fermi surface is small along the antinodal direction. This has consequences that can be better understood by rewriting the Hamiltonian in $\mathbf{k}$-space:

$$H_{JT} = \lambda \sum_{a,a',\mathbf{k},\mathbf{q},\gamma} \delta_{\gamma,a,a'} C_\mathbf{k}^{\dagger a} C_{\mathbf{k}-\mathbf{q}}^{a'} Q_{\gamma,\mathbf{q}}. \quad (7)$$

Switching to momentum space remarks how the distortions break the translation invariance of the Hamiltonian. This portion of the Hamiltonian, $H_{JT}$, will be here considered as a perturbation. Let us label the eigen-states of the unperturbed Hamiltonian as $|k\gamma\rangle$, where $k$ is the momentum and $\gamma$ refers to a particular linear combination of $e_g$ orbitals. For non-degenerate states, the first order correction is $\langle k\gamma|\mathrm{H}_{JT}|k\gamma\rangle$. This is only-non-zero for $\mathbf{q}$=0, since otherwise it mixes different eigenstates. The $\mathbf{q}$=0 modes correspond to changes in the shape of the lattice. Indeed real systems present such distortions, that lead, for example, to tetrahedral unit cells. However, they are relatively small, and we are not interested in studying them here, so our boundary conditions for distortions impose a cubic lattice and fixed lattice parameters. More interesting for our purposes is applying perturbation theory to degenerate states, by diagonalizing Eq. (7) within each subspace of degenerate eigenstates. In the $|k\gamma\rangle$ basis, the Hamiltonian has only off-diagonal terms and therefore, the sum of energy shifts within each subspace is zero. Since the splitting is proportional to the off diagonal terms, only subspaces with energies close to the Fermi energy, where occupations might change, need to be considered.

For small perturbations, the nesting vector needs to *connect* states very close to the Fermi energy. But the criteria to consider degeneracy within subspaces is, of course, that the energy differences are small compared to the terms in the perturbing Hamiltonian (7). We argue here that since electron-phonon coupling in manganites is not small, even non-perfect nesting near the Fermi surface leads to lattice distortions and a suppression of the spectral weight in the antinodal direction. Entropy also plays an important role, favoring distortions. Our numerical results indicate that the well studied coupling of the electronic $e_g$ orbitals to the Jahn-Teller lattice distortions (Eq. (A.3)) is responsible for the phenomena observed in experiments: a pseudogap phase with a gapped dispersion relation along the antinodal direction and coherent spectral weight along the nodal direction, with the peculiar Fermi surface forming the, so called, Fermi arcs.

Since couplings are not small, we go beyond perturbation theory and use Monte Carlo numerical simulations to study the model in Eq. (1). The undistorted system involves the calculation of only one configuration, therefore it is possible to go to large sizes or include many phases in the twisted boundary conditions, and track the nesting vectors of the Fermi surface as a function of doping. The nesting vector varies continuously with doping, and for $x=0.51$ it is very close to $(\pi/2, 0)$, so that this vector connects the two points in the antinodal direction, and provides non-perfect nesting in the sense discussed above. Wave vectors with $k_x = \pi/2$ and small $k_y$ components are also good candidates, and so are their equivalents by symmetry. A $\pi/2$ wave vector component corresponds to a wave length of 4 lattice sites, which is commensurate with lattice sites accessible to the computationally expensive Monte Carlo simulations. The situation is different for $x=0.4$ and $x=0.6$ (Section V), since the nesting vector is incommensurate with any of the lattice sizes reachable by Monte Carlo simulations. Anyhow, we have performed calculations for different values of $\lambda$ and $T$ for both $x=0.51$ and $x=0.5$, and they are discussed in Section IV. We find that there is hardly any difference between the two dopings. That is again in agreement to the proposed physical picture of nearly perfect nesting. Not the exact value of the nesting wave vector, but the fact that there are nearly parallel segments of the Fermi surface, is what determines the properties of the system. We will show below that near perfect nesting and large electron-phonon coupling are responsible for the peculiar Fermi surface of small-bandwidth manganites in the FM phase.

Figure 3 shows both the band diagram and the Fermi surface for $x=0.5$ projected over the two $e_g$ orbitals. Besides the curvature, there is also a clear difference between the nodal and antinodal directions. While the nested parts along the antinodal direction have almost equal weight for the two orbitals, in the nodal direction the Fermi surface is purely of $x^2 - y^2$ character. This lack of mixing along the nodal direction is due to topological reasons, and it will be there for any system with orbitals of the same symmetry on a square lattice.

## IV. $x=0.5$

### A. Low temperature

The total density of states shown in Fig. 4 gives a good idea of the different electronic states in manganites. For small electron-phonon coupling ($\lambda = 1.3$ or smaller) the system is a metal. The density of states is fairly smooth, since the Van Hove singularities are smoothed out by temperature. If we increase the electron-phonon coupling by less than 10% ($\lambda = 1.39$), the density of states at the Fermi energy drops by approximately 50%. This is compatible with the pseudogap phase that shows up in experiments, and has been characterized in detail recently for bilayer manganites. Further increases in $\lambda$ diminish the density of states further, and already for $\lambda = 1.7$ there is clear gap. The situation is summarized in Fig. 5. These large changes in electronic structure with relatively small changes in parameters is well known for manganites. A reduction on the density of states at the Fermi energy has also been reported in one-dimensional[17] and in one-orbital models[17,24]. Considering a more realistic model allows us to make a direct comparison with experimental results. For the rest of this Section, and the next Sections dealing with other dopings, we will focus on this interesting pseudogap phase, its physical origin, and the similarities with cuprates.

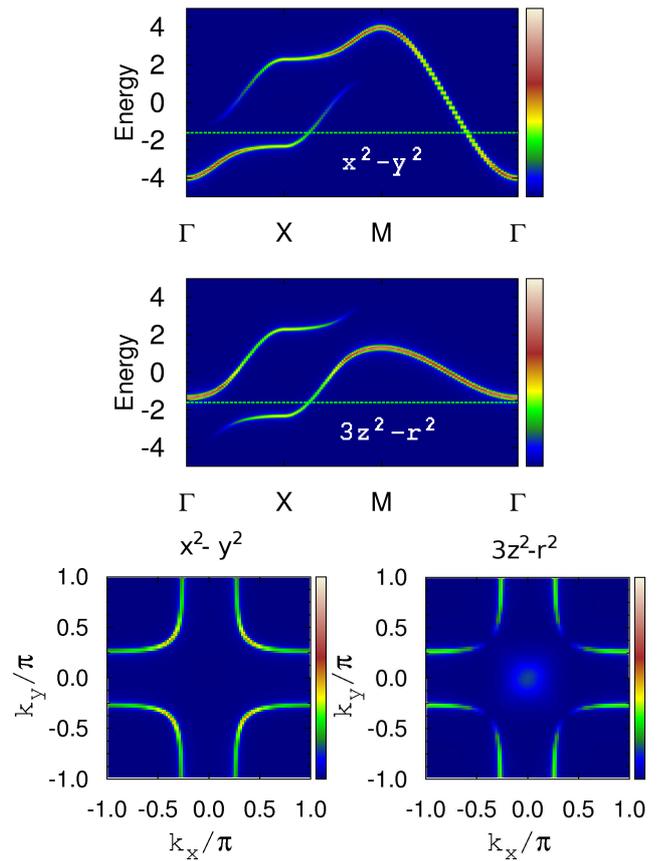

FIG. 3: Bands (top two panels), and Fermi surface (bottom) projected over the two $e_g$ orbitals, as indicated. The higher overlap of neighboring $x^2 - y^2$ orbitals within the plane leads to the *larger bandwidth* of the corresponding projection. Notice how the symmetry of the orbitals prevents mixing in the (1,1) direction (M-$\Gamma$). Consequently, in that direction, all spectral weight of the Fermi surfaces in Fig. 2 comes from the $x^2 - y^2$ orbital, except for the small electron pocket for $x=0.4$ (light gray in Fig. 2) which is almost completely of $3z^2 - r^2$ character. The spectral weight near the Brillouin zone face, on the contrary, comes from both orbitals.

The physical picture arising from our calculations is illustrated by the numerical results presented in Fig. 6. There it can be seen that distortions with certain wave vectors are much more likely to take place. As shown in the inset, and explained in Section III, these wave vectors are the ones that connect points of the Fermi surface (or, more precisely, points in the bare dispersion relation with an energy close to the Fermi energy).

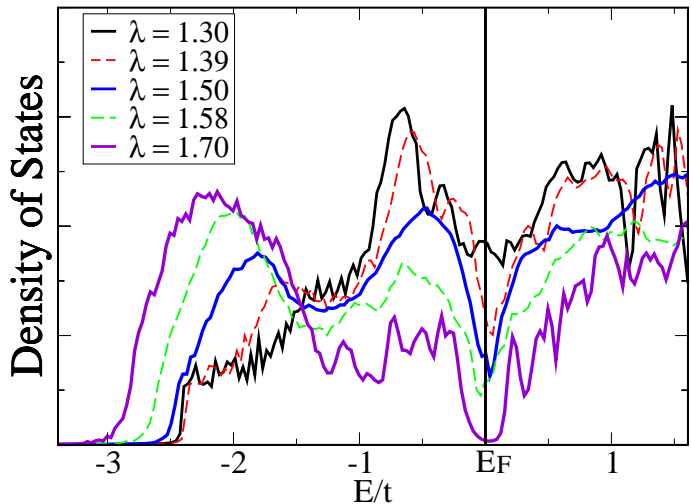

FIG. 4: Total density of states for $x=0.5$ and different electron-phonon couplings ($\lambda$'s) from Monte Carlo simulations at $T=0.01$. The Fermi energy ($E_F$) for the different parameters is taken as zero energy. As $\lambda$ increases, the density of states at the Fermi energy decreases, and, for a large enough $\lambda$, a gap opens. Experiments[1,3,13] and the present work indicate that layered manganites are in the pseudo-gap regime with a strong reduction of the density of states at the Fermi energy. A small imaginary part has been given to each eigenvalue arising from the Monte Carlo simulations in order to obtain a smooth curve.

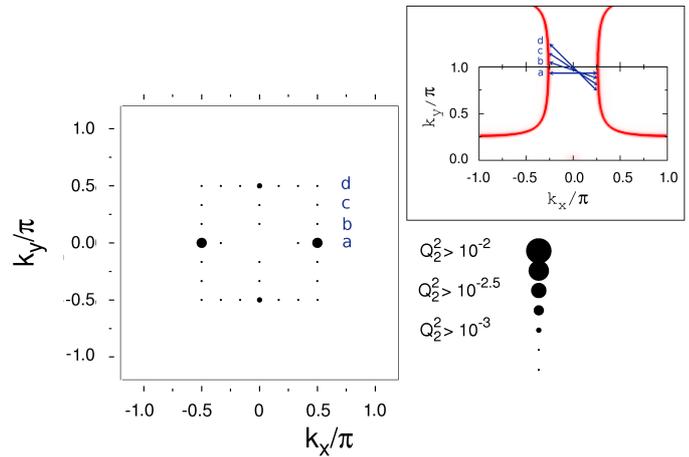

FIG. 6: Monte Carlo average of thermally excited distortions, as a function of their wave vector. For each $(k_x, k_y)$ point, the size of the dot is proportional to the thermal average of $Q_2^2(k_x, k_y)$ (see text). The main graph shows the thermal average of the $Q_2$ mode, for $\lambda=1.39$ and $T=0.01$. The inset at the top-right shows the Fermi surface of the undistorted system, where the wave vectors labeled in the main panel are shown connecting some points of the Fermi surface. For the thermal averaged Fermi surface with these parameters, see Fig. 7. The finite size of the simulation lattice limits the allowed wave vectors of the distortions, and the number of them is the same as the number of sites (144 unless otherwise noted).

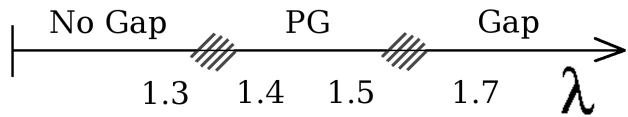

FIG. 5: Cartoon characterizing the different electronic states of the system, as a function of $\lambda$ at low $T$. As the values at which different electronic structures are found depend on doping and temperature, no precise boundary for them is shown. The values of $\lambda$ indicated correspond to simulations with $x=0.5$, and $T=0.01$.

In the numerical simulations, the possible wave vectors of the distortions are limited by the size of the lattice. As shown in Fig. 6, this is not very important for the doping $x=0.5$ we discuss in this Section. The most favorable distortions have wave vectors with either $x$ or $y$ components of $\approx 1.7$, close to the $\pi/2$ allowed by the lattice. For other dopings this might not be the case. In those situations, the simulation still chooses the most favorable distortions allowed by the boundary conditions. A larger size will probably mean a closer wave vector to the ideal one. Therefore, the physical phenomena discussed here might appear for electron-phonon couplings that become slightly smaller as we move to larger lattice sizes. Anyhow, our simulations show that this effect is small.

When distortions appear, the system looses the translation invariance symmetry, and the spectral function is no longer zero or a constant. Fig. 7 shows these changes in the spectral function, for the same parameters as in Fig. 6. We see how a reduction of the spectral function takes place near the antinodal direction. Obviously, this is the same reduction in the density of states near the Fermi energy seen in Fig. 4 for $\lambda = 1.39$; in **k**-space, the reduction takes place in the *nearly nested* parts of the Fermi surface.

As we increase $\lambda$, this effect is more pronounced. For $\lambda=1.5$, there is a clear gap when we look along the antinodal direction (Fig. 8). The band along the nodal direction is still well defined, despite a small broadening. This enhances the arc-like aspect of the Fermi surface shown in Figs. 8 and 9. An effect already considered in experiments becomes clear when comparing these figures. Although increasing $\lambda$ from 1.5 to 1.58 results in a reduction of the spectral weight both in the nodal and antinodal directions, this reduction is not easy to see when the intensity scale is set by the maximum in $A(k, E_F)$. As a result, Figs. 8 and 9, which we have chosen to plot with different scales for easier comparison with experiments, look alike except for a larger broadening of the Fermi arc for larger $\lambda$. A similar effect has been discussed in Ref. 25 regarding the evolution of the Fermi arcs with doping.

The distortions follow the general pattern discussed for the $\lambda = 1.39$ simulation. In Fig. 8 we see the $Q_2$, (as in Fig. 6), and $Q_1$ modes for $\lambda = 1.5$, and the same modes are shown for $\lambda = 1.58$ in Fig. 9. The $Q_2$ mode follows much closer the nested wave vectors, while $Q_1$ is more influenced by thermal noise. This is not surprising as $Q_2$ couples the two $e_g$ orbitals (see Eq. (3)), and we already discussed that the part of the Fermi arcs that is nested has weight from both $x^2 - y^2$ and

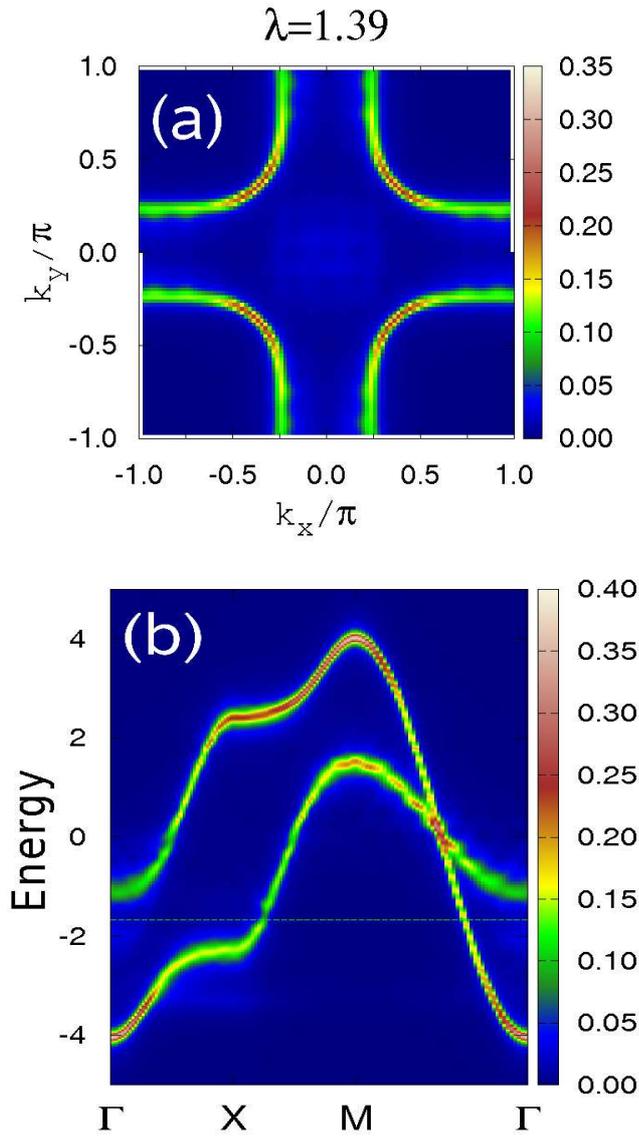

FIG. 7: Spectral function arising from the Monte Carlo simulations. The same parameters as in Fig. 6 are used ($\lambda$=1.39, $T$=0.01, and $x$=0.5). (a) Energy cut at the Fermi energy, and (b) A(**k**,E) for **k** along the high symmetry directions. Already for this coupling the distortions shown in Fig. 6 result in a reduction in the spectral weight near the Brillouin zone face (antinodal direction), that can also be observed in both projections.

$3z^2 - r^2$.

For $\lambda \gtrsim 1.6$ an ordered phase appears at $T$=0.01. The fourfold rotation symmetry of the system is broken and the simulations show an ordered pattern of distortions, coupled to orbital order, with $Q_2(\pi/2,\pi/2)=Q_2(-\pi/2,-\pi/2)\sim 1$, and $Q_2(\pi/2,-\pi/2)=Q_2(-\pi/2,\pi/2)=0$. This phase was already discussed in Ref. 26. The orbital and lattice order is similar to the CE phase, and, more interestingly, also shows a charge order with double periodicity, $(\pi,\pi)$, that couples to the $Q_1$ mode.

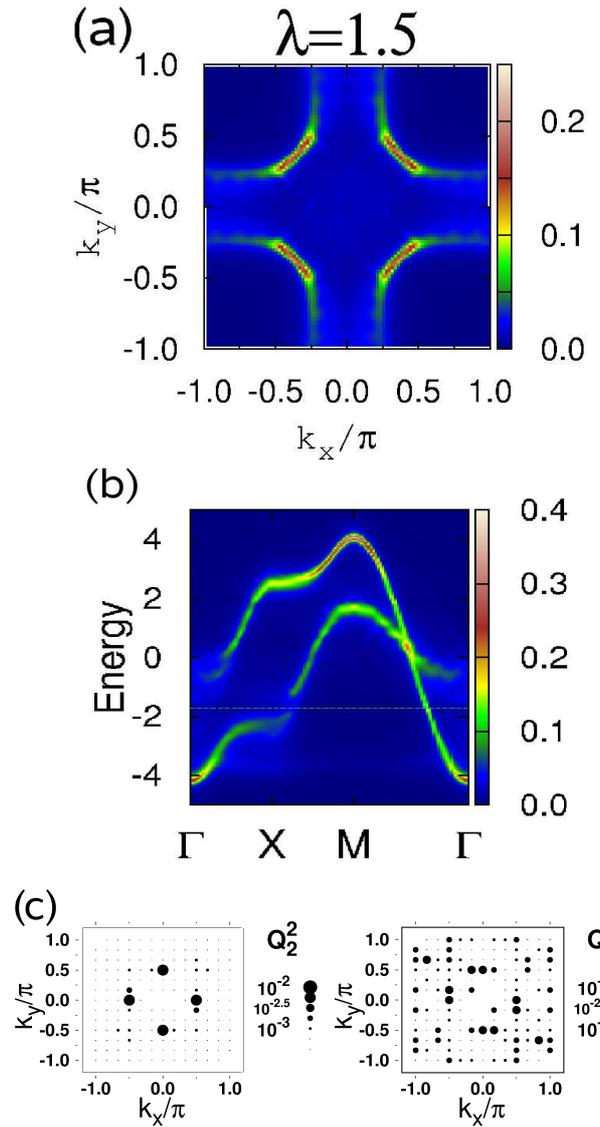

FIG. 8: Results of the Monte Carlo simulation for $\lambda$=1.5, $T$=0.01, and $x$=0.5, using a 12×12 lattice. The spectral function cut at the Fermi energy is shown in (a), and its value along the high symmetry directions, in (b). For these parameters, a gap is already formed along the antinodal direction (X-M), while in the nodal (M-$\Gamma$) direction there is a coherent band. This turns into a Fermi-arc-like Fermi surface, with hardly any spectral weight left near the Brillouin zone face, and a well-defined coherent peak near the zone diagonal (pseudogap regime). (c) shows the $Q_1$ and $Q_2$ (see text) distortions in **k**-space. A larger electron-phonon coupling as compared to Fig.6 make distortions with more wave vectors noticeable under this scale (same as Fig. 6). However, the distortions with the appropriate nesting wave vectors are more favored by $\lambda$, and they produce the arc-like Fermi surface shown if (b).

Two aspects of the ordered phase appear in the pseudogap phase as $\lambda$ approaches the transition value. In Fig. 9, the wave vectors for which the $Q_2$ is significant have already started to loose the fourfold rotation symmetry. Probably a long enough Monte Carlo simulation will restore it by visiting the two sym-metric possibilities, but the proximity to the phase transition makes the Monte Carlo dynamics slower, and we are unable to confirm this. Our results also indicate that when the thermal average of the $Q_2$ modes with $k_x$(or $k_y$)=$\pm\pi/2$ is large enough the $Q_1$ mode with double periodicity is also substantial (Figs. 8 and 9). Similarly to the ordered phase, orbital correlations induce charge correlations with half the period in the symmetric pseudogap phase.

It is also interesting to note that the gap in the ordered phase has a minimum in one of the diagonal directions (the rotation symmetry is lost) of the unfolded Brillouin zone, that is, the nodal direction where the qausiparticle peaks are found in the pseudogap phase. These facts are compatible with fluctuations induced by the proximity to the ordered phase as a physical origin of the pseudogap phase, at least for $\lambda = 1.58$ and $x$=0.5. However, the results for other dopings presented in the next sections indicate that the pseudogap phase is induced by non-perfect nesting and intermediate to large electron-phonon coupling, and it is independent of the existence of a true long-ranged ordered phase.

Although the $Q_2$ distortions follow more closely the nested wave vectors than $Q_1$, those do not suffice to reproduce all the experimental results. It is possible to explore the role of $Q_1$ by slightly modifying the model Hamiltonian. $Q_1$ distortions can be reduced by increasing the elastic cost of this mode, with a factor $\beta >1$ multiplying the $Q_1^2$ term in Eq. (3). We have run simulations with $\beta >1$ and found that this suppresses or reduces the range of stability of the pseudogap phase, favoring the metallic and the ordered phase. In particular, for $\beta$=2 and small $\lambda$ ($\lesssim$1.39) the lost of spectral weight is smaller than for $\beta$=1. And for $\lambda$=1.45 the system is already in the ordered state. The fragility of the pseudogap phase with $\beta$ can be understood by defining a Jahn Teller angle $\theta$ by $Q_1$=$Q\sin\theta$ and $Q_2$=$Q\cos\theta$. An increase of $\beta$ favors $\theta$=0 and $\theta$=$\pi$, which leads more easily to an ordered phase, and penalizes configurations with other values of $\theta$. Since in the pseudogap phase $Q_2$ follows the nested wave vectors more closely than $Q_1$, $\theta$ is small for the nearly nested $k$'s, (certainly smaller than $\pi/4$), but non-zero. When $\theta$ is constrained to be 0 or $\pi$, the pseudogap phase disappears from the phase diagram within the precision of our calculations.

In Fig. 10 the dispersion relation and the variation of the quasi particle (QP) weight with lambda along the nodal direction is illustrated. They have been determined by fitting the momentum distribution curves (MDC, in the ARPES literature notation). As explained in section II.C, the twisted boundary conditions improve the determination in the momenta of electrons. With the 8x8 system size and 12x12 twisted boundary conditions used for Fig 10, the number of momenta coming out of the simulations is still insufficient for a detailed analysis of the dispersion relation. We therefore follow the standard fitting procedure. For each energy, the spectral weight is fitted to Lorentzian functions around the maxima, the center of these Lorentzian's define the dispersion relation, while their amplitude determines the quasi particle weight. Figure 10shows that both the dispersion relation, and the quasi particle weight, for $\lambda = 1.39$ are very close to the $\lambda = 0$ curves (dotted line in Fig. 10). This is in sharp contrast with the clear reduction of spectral weight that already takes place for this $\lambda$ in the antinodal direction (Fig. 7 (a) and (b)). When lambda is further increased, an *s* shape feature develops in the dispersion relation. This has been already observed in ARPES experiments[1,5]. This peculiar shape is accompanied by a marked reduction of the spectral weight, that for these values of $\lambda$ also starts to occur also in the nodal direction. Our numerical accuracy is not enough to determine the effective mass for the higher (and more interesting) $\lambda$'s, as this accounts to numerically calculate a second derivative of a curve with rapidly changing curvature. However important conclusions can be drawn from the results in Fig. 10. It is doubtful that transport properties can be predicted by only measuring the effective mass along the nodal direction, as the spectral weight and curvature (effective mass) depend strongly on the k point on the Fermi surface. Not only transport properties, as already suggested in Ref. 1, for the couplings examined here, extreme care should be taken when deriving physical quantities from dispersion relation curves. Eliashberg relation between the electron-phonon coupling and effective mass(m*), $\lambda$=[(m*/$m_o$)-1], appear to fail in our situation, it fails to provide a rough estimate of the couplings. Within the precision of our calculation the effective mass is equal for $\lambda = 0$ and $\lambda = 1.39$, instead of a factor of 2.8 as predicted by such a formula. Needless to say, the tensorial character of the effective mass is more important in these cases where the anisotropy of the Fermi surface is enhanced by interactions.

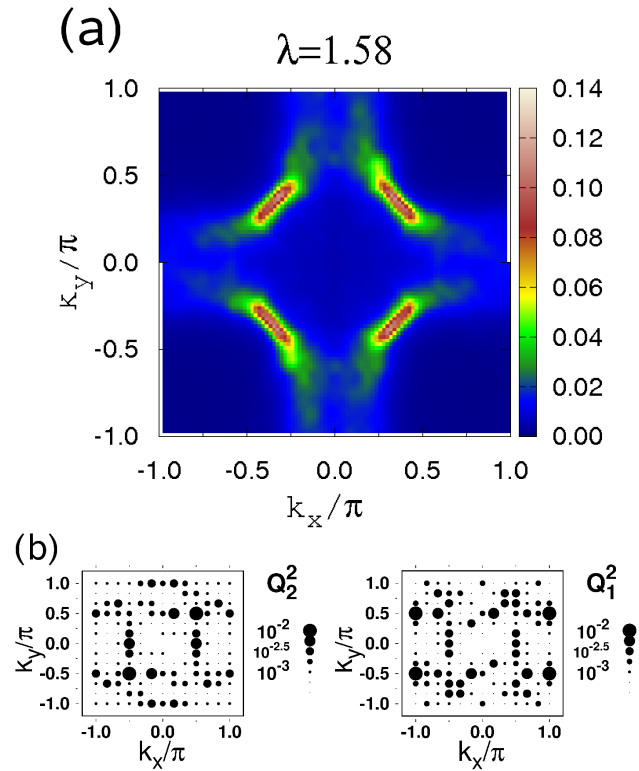

FIG. 9: Monte Carlo simulation results for $\lambda$=1.58, $T$=0.01, and $x$=0.5, using a 12×12 lattice: (a) A(**k**,$E_F$); (b) $Q_1$ and $Q_2$ distortions in **k**-space. The system is close to opening a complete gap: spectral weight at the Fermi energy survives only near the (0,0) to ($\pi$,$\pi$) diagonal, or nodal direction. As $\lambda$ increases, distortions with more wave vectors appear on the system (see Fig. 8). Notice that if $Q_2$ induces orbital correlations with period 4a (wave vector $\pi/2a$), charge correlations with period 2a can be expected that favor $Q_1$ modes also with period 2a (wave vector $\pi/a$).

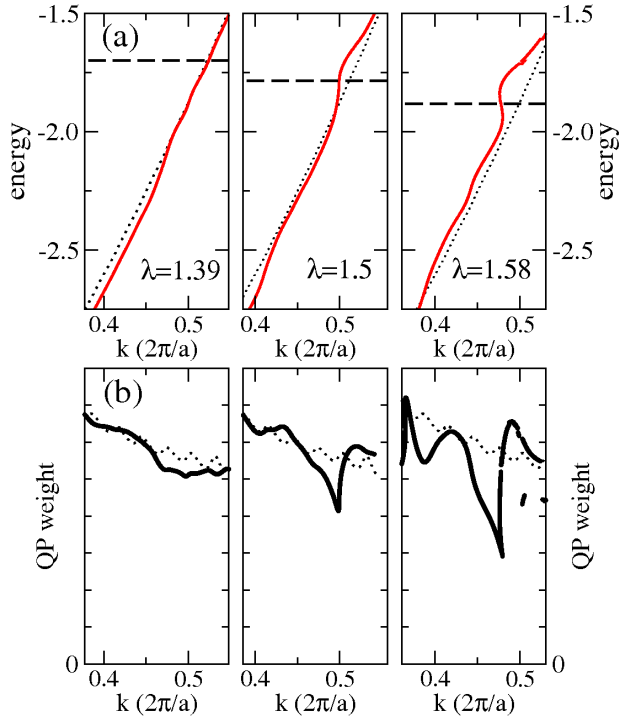

FIG. 10: Dispersion relation (a) and quasi particle (QP) weight (b) along the nodal direction, for different $\lambda$'s. The dashed lines indicate the position of the Fermi energy for the different couplings. To overcome momentum quantization, a standard fitting procedure of momentum distribution curves has been followed (see text). This induces a small oscillatory error in the determination of the QP weight. The $\lambda = 0$ (dotted lines) curves, determined with the same procedure has been included to illustrate the error and for comparison purposes. Numerical accuracy is not enough to determine the effective mass for large $\lambda$'s. Curvature rapidly changes around $k_F$, where an $s$ shape dispersion relation is found, similar to the one reported in experiments[1,5]. As spectral weight and curvature change along the Fermi surface we don't expect this curvature to directly determine transport properties. QP weight at $k_F$ is lost as $\lambda$ increases and the system develops a pseudogap, but this takes place at larger $\lambda$'s in the nodal direction, as compared to the antinodal direction (see Fig. 7 (a) and (b)).

### B. Effect of Temperature

All the results of the Monte Carlo simulations presented above correspond to different values of the electron-phonon coupling and a particular temperature T=0.01. The effects of changin the temperature are discussed next, and summarized in Fig.11 It is remarkable that, for some couplings, the system appears essentially undistorted in the simulations at very low T, and develops a pseudogap due to the correlations within the thermal induced distortions. This is the case of lambda=1.4 discussed below. An open issue is whether the pseudogap can survive to T=0. Quantum fluctuations of the phonons (not considered here) might work in a similar way as thermal fluctuations do for the temperatures considered in this work.

Figure 12 presents the evolution of distortions with temperature for a particular electron-phonon coupling, $\lambda$=1.4. We still focus on the influence of the lattice distortions on the electronic structure keeping the magnetic degree of freedom frozen. This coupling strength is particularly interesting, since the ordered phase is not yet stable, and the system appears essentially undistorted at very low temperatures. However, $\lambda$ is big enough to induce strong fluctuations with temperature, that in turn qualitatively change the electronic spectral function. Results for both 8×8 (left column) and 12×12 (right column) system sizes are included to illustrate the issue of possible finite size effects.

For $T$=0.08, Fig. 12 shows the pattern of distortions discussed in previous sections, with the $Q_2$ mode active when **k** corresponds to the nested wave vectors. The absolute value is not large enough to significantly alter the Fermi surface in Fig. 13, where only a small reduction of spectral weight as compared to the undistorted system is observed. As the temperature increases, distortions with all wave vectors are thermally excited, but the increase with temperature is larger for distortions with the nesting wave vector. For $T$=0.015 in Fig. 13, there is clearly less spectral weight along the antinodal direction as compared to the nodal direction.

For the larger temperatures in Figs. 12 and 13 magnetic excitations might be important ($T$=0.01$\simeq T$=60K). We expect them to enhance the effect of lattice distortions. Magnetic excitations reduce the mean value of the kinetic energy as described by the double exchange model,[18] thus increasing the relative importance of the electron-phonon coupling.

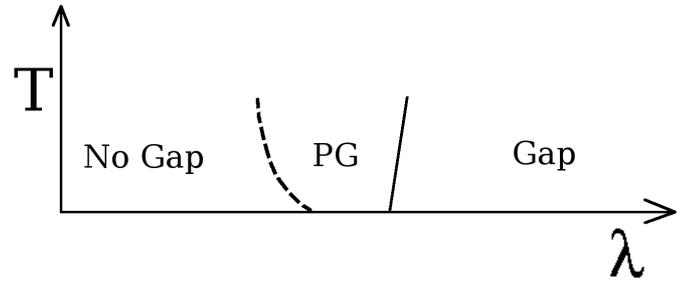

FIG. 11: Schematic representation of the evolution of the different electronic states as a function of temperature. The transformation from a normal metal with smooth density of states at the Fermi energy to the Pseudogap (PG) state, with a minimum of the density of states at the Fermi energy, is continuous (see Fig. 4). Our results indicate that temperature favors the pseudogap, as explained in the text. For $x$=0.5, the phase with a gap is an ordered phase and the continuous line separating the PG and the ordered phase corresponds to a first order transition.

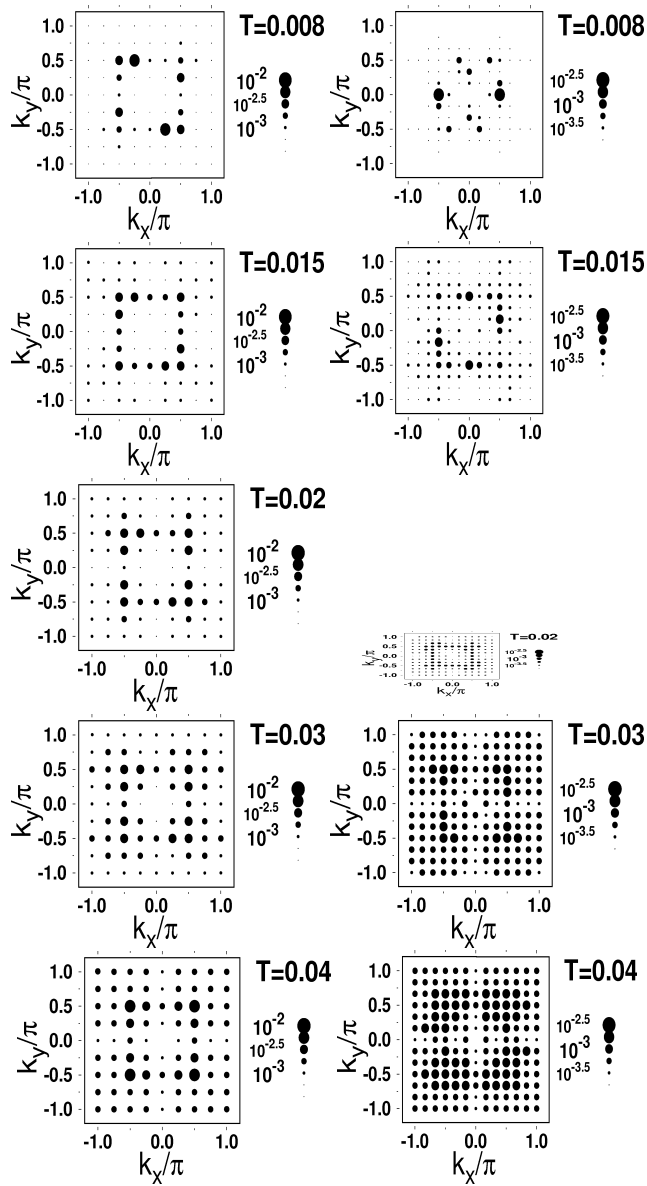

FIG. 12: Evolution of the $Q_2$ Jahn Teller distortions in **k**-space as a function of temperature for an $8\times 8$ (left column) and $12\times 12$ (right column) lattices, for $\lambda = 1.4$, at half doping. In general, distortions are overestimated at low temperatures for small lattice sizes (specially at temperatures lower than the ones shown), see legend. Finite size effects are small, and the square-like feature responsible for the Fermi arcs consistently appears for these sizes and temperature range.

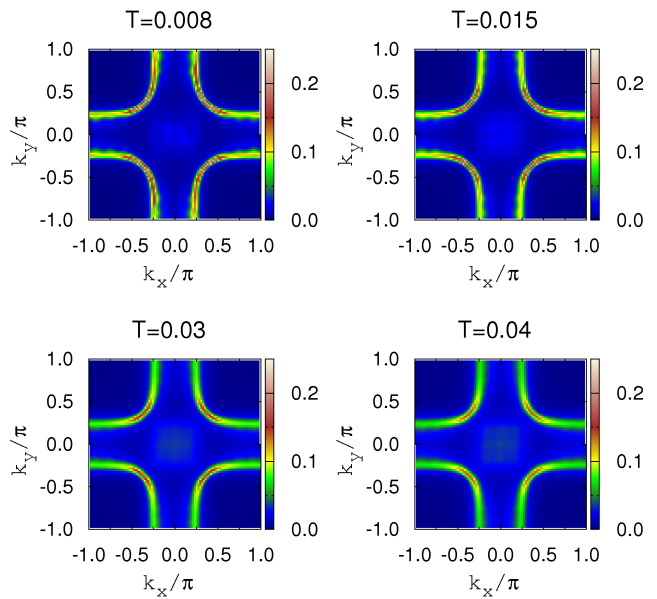

FIG. 13: Spectral function at the Fermi energy for some of the temperatures in Fig. 12 ($\lambda = 1.4$, $x=0.5$). At low temperatures the system is undistorted for this $\lambda$. The nearly perfect nesting of the antinodal portions of the Fermi surface makes the distortions with the proper nesting vector much more likely to be thermally excited, producing the Fermi arcs. The color scale is the same for all temperatures. At higher temperatures, the entropy contribution to the free energy dominates, and the spectral weight along the nodal and antinodal directions becomes similar.

The absence of important finite size effects can also be observed in Fig. 12. We have compared distortions for $8\times 8$ and $12\times 12$ lattices, for the same $\lambda$'s and temperatures. As expected, small lattices enhance correlations, leading to slightly larger lattice distortions (note the different scales). At very low temperatures, this enhancement of correlations leads to the appearance of an ordered phase for the $8\times 8$ cluster at $\lambda = 1.4$. However, the Monte Carlo simulations using the $12\times 12$ system already recover the physically meaningful undistorted result. This has been confirmed by direct optimization of the oxygen position in systems as large as $30\times 30$ lattice sites.

Experiments show that increasing the temperature results in an enhancement of fluctuations, as it has been observed by Raman experiments.[27] This can be understood in the light of a double exchange model. A temperature rise results in a reduced magnetization, a smaller hopping due to double exchange, and a reduction in the bandwidth, making the effective coupling, measured by the ratio of Jahn-Teller energy to electronic bandwidth, larger. The results in this section indicate that also entropy favors lattice fluctuations, and that these lattice fluctuations are not random, but have, for a certain range of temperatures, correlations induced by the electronic structure.

## V.  OTHER DOPINGS: $x=0.4$ AND $x=0.6$

With the insight gained by the study of the $x=0.5$ case now we examine the simulations for $x=0.4$ and $x=0.6$. This is important for two reasons. One is that the real layered systems become CE-antiferromagnetic at $x=0.5$. Although this FM to AF transition is well understood,[21] and it does not affect our arguments, calculations with $x=0.4$ should facilitate the comparison with experiments. $x=0.6$ is similar to $x=0.5$ with an ordered phase for a narrow doping range, but our calculations is nested for the undistorted system (Fig. 2). Our simulations indicate that this causes the corresponding distortions (Fig. 14), to be excited at lower temperatures. For $\lambda = 1.35$, the distortions are small, but as $\lambda$ increases, distortions with wave vectors connecting electronic eigen-states in **k**-space with energy close to the Fermi energy are excited. Figure 14 also shows the Fermi surface resulting from those distortions. Besides being somewhat more noisy due to slower Monte Carlo dynamics, the resulting Fermi arcs are shorter, as the nested portion looses spectral weight. For $\lambda = 1.6$ the Fermi surface has almost disappeared (notice the different scale for different $\lambda$'s in the Fermi surfaces of Fig. 14). The spectral weight left is once again very similar to what has been observed in cuprates with four quasiparticle peaks in the four nodal directions.

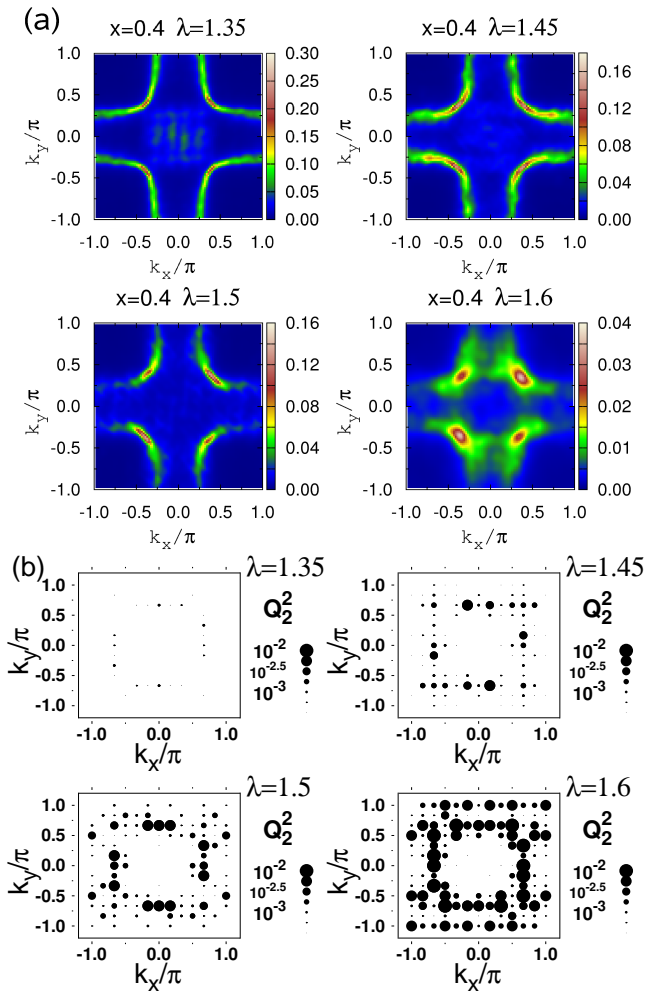

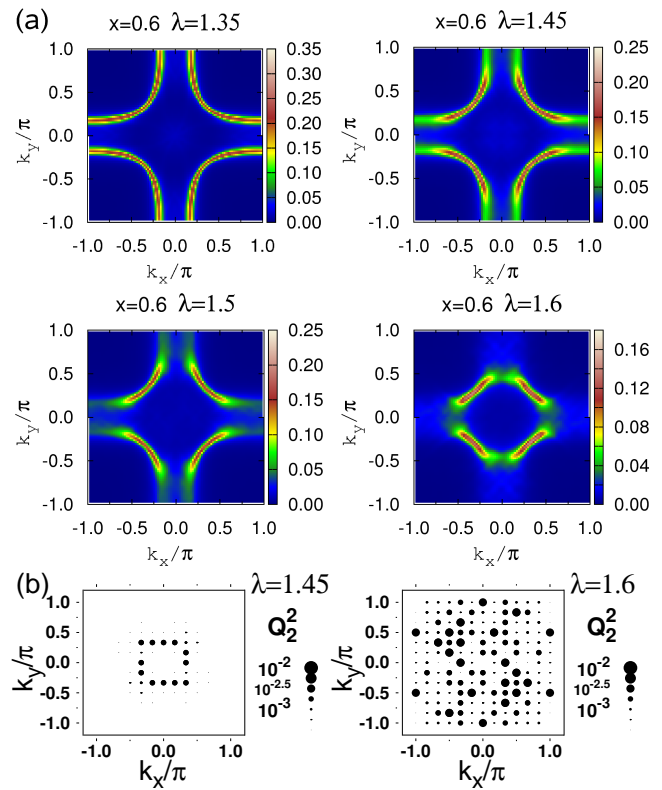

FIG. 14: Evolution of the system as a function of $\lambda$ for $T=0.001$, $x=0.4$, using a $12\times 12$ lattice. (a) shows Fermi surface changes. Both the detailed way in which the Fermi surface changes, and the temperature at which the changes take place depend on doping. For this temperature and the lower two values of lambda, the system is essentially undistorted at half doping. Longer nested portions of the Fermi surface (see Fig. 2) for $x=0.4$ result in smaller coherent quasi-particle peaks at lower temperatures. As $\lambda$ grows, the Fermi surface shrinks to small Fermi arcs, making different dopings much more alike (compare $\lambda = 1.5$, and $\lambda = 1.6$ with Figs. 8, 9, and 15). $Q_2$ Jahn Teller distortions in **k**-space are shown in (b). The behavior is similar to half-doping, but for this doping the nesting wave vectors cannot be well approximated with a wave vector commensurate with the lattice. For low values of $\lambda$, the distortions with the nearest commensurate wave vector components, $2/3\pi/a$, are favored, and as $\lambda$ increases, distortions with the next nearest one, $\pi/2a$ are also present in the system.

FIG. 15: Monte Carlo simulations results as a function of $\lambda$ for $T=0.01$, $x=0.6$, using a $12\times 12$ lattice: Spectral function at the Fermi level is shown in (a), and lattice distortions in (b). For this doping, larger $\lambda$'s are needed to achieve the reduction in the spectral weight along the antinodal direction, and the $\lambda=1.35$ spectral function appears essentially undistorted. Notice how $\lambda=1.6$ is large enough such that distortions can not be understood within the perturbation theory arguments presented in the text. Like the cuprates, in the strong coupling regime, the Fermi surface is very similar for dopings as different as 10%-20% (See Figs. 9 and 14)

should present general trends observable at dopings close to the ones presented here. Examining different dopings is also important in order to confirm that excited distortions follow the changes in the nesting vectors with doping, providing another evidence for the physical picture presented in previous sections.

For $x=0.4$, there is a larger portion of the Fermi surface that

The doping case $x=0.6$ follows the general trend explained in this article, with only small differences. The wave vectors of the significant distortions in Fig. 15 are smaller, as compared to $x=0.5$ and $x=0.4$, as expected from the undistorted Fermi surface (Fig. 2). Notice, however, that for $\lambda=1.35$ the spectral function is very similar to the undistorted one, and correlations between distortions (not shown) do not follow the nested parts of the Fermi surface. This might be caused by the fact that the nested parts of the Fermi surface are smaller. $\lambda$ needs to become larger as compared to other dopings, and only for $\lambda=1.45$ we do recover a picture consistent with other dopings (Fig. 15) with the same square like features as $x=0.5$ and $x=0.4$. As the nested potions of the Fermi surface are smaller, we observe longer Fermi arcs for every calculated coupling. Notice how, even for $\lambda=1.6$ where the arguments of perturbation theory seem to be no longer valid, apparently noisy and disordered distortions (Fig. 15 (b), $\lambda = 1.6$) result in a suppression of the spectral weight in the antinodal direction.

As recently pointed out by Mannella,[28] there is a correlation between the Fermi surfaces obtained by ARPES and transport properties. The larger spectral weight at the Fermi energy for dopings around $x=0.6$ correlates with a larger in-plane conductivity[29] as compared to other dopings.

Similarly to the case of $x=0.5$ doping, we need to go to higher temperatures, as compared to the $x=0.4$ doping, to observe same magnitude lattice distortions in the simulations. Notice that $x=0.4$ is the only doping at which a complete suppression of coherent spectral weight in the antinodal direction has been observed in experiments.

## VI. CONCLUSIONS

This article presents a detailed study of the evolution of the spectral function in a model for manganites as a function of temperature and electron-phonon coupling; several hole dopings were examined. This large effort was made possible by the use of twisted boundary conditions, and the reduced computational effort that results in focusing on the relevant degrees of freedom. Apart for the double exchange mechanism, the same interactions and similar approximations had been previously taken into account to qualitatively explain a temperature induced metal insulator transition[14]. This work, thus, provides an approximation to the physics of manganites compatible with previous ones. With this framework, the very remarkable experimental observation of the existence of Fermi arcs[1] in manganites was here addressed.

Monte Carlo numerical simulations show a strong reduction or suppression of the spectral weight in particular parts of the Fermi surface, in agreement with experimental results[1,3] and similarly to the pseudo-gap phase of high-$T_c$ cuprates. The microscopic origin is in the cooperative Jahn-Teller distortions. Nesting arguments show that as the wave vector of these correlated distortions is such that it connects different parts of the Fermi surface, it can lower the electronic energy. Our numerical results support similar previously proposed scenarios.[3] For the range of couplings relevant to the experimental results, states with energies close to the Fermi energy also play a role, and distortions with several wave vectors appear in the simulations. We argue that this non-perfect nesting effect is due to the significant electron-phonon coupling and the effect of temperature. Furthermore, these correlations are found even for values of the coupling where the system is essentially undistorted at low temperatures, which highlights the role of fluctuations. A natural extension of this work is therefore the study of a system with phononic quantum fluctuations.

Different physical regimes were studied here, with particular attention to the evolution of the spectral function as the electron-phonon coupling was varied. This is the only parameter not directly measurable in experiments (as opposed to temperature and hole doping). Extensive numerical calculations show that this effect is not a result of fine tunning of parameters, but robust with respect to changes in temperature and doping. The possibility to continuously tune the electron-phonon coupling allow us to explore the high sensitivity of the system to changes in this parameter. We propose that this sensitivity is behind the different experimental results about the complete suppression[1] or strong reduction[3] of spectral weight in the antinodal direction for the $x=0.4$ doping. Different samples, different amounts of disorder, or slight changes in the composition can lead to small changes in the effective coupling that are amplified by the system's response.

Changes with temperature are also explored. Keeping the spins frozen in the ferromagnetic state does not take into account the kinetic energy reduction of the carriers, but allow us to observe how the thermally induced fluctuations of the lattice distortions reduce the spectral weight at the Fermi energy. This reduction takes place preferentially in the antinodal directions, and would *cooperate* with a larger Jahn Teller to kinetic energy ratio arising from double exchange to lead to the observed large changes in the spectral function with temperature[4].

## VII. ACKNOWLEDGEMENT

We are grateful to Maria Daghofer and Norman Mannella for helpful discussions. This work was supported by the NSF (DMR-0706020) and the Division of Materials Science and Engineering, U.S. DOE, under contract with UT-Batelle, LLC.

---

## APPENDIX: ELECTRON-PHONON HAMILTONIAN

In several previous efforts, the sign of the coupling between the breathing mode and the density was positive. We will briefly argue why it should be negative (as in Eq. (3)) and the small difference that might arise for the two different signs.

First, let us write down the value of the $Q$ modes in terms of the oxygen displacements. Consider a Mn ion $i$: both in the three dimensional perovskite structure and in layered manganites, it has six surrounding oxygen ions. We label $i, x$ and $i - x, x$ the neighboring oxygen ion in the positive and negative $x$ directions. It is enough to consider oxygen displacements along the Mn-O direction, so let $\mathbf{u}_{i,x}$ and $\mathbf{u}_{i-x,x}$ be the displacement of these oxygens along the $x$ direction, and $Q_x$ the change in distance between them along the $x$ direction, $Q_x = \mathbf{u}_{i,+x} - \mathbf{u}_{i,-x}$. $Q_y$ and $Q_z$ can be similarly defined. Since each oxygen is nearest neighbour to two Manganese ions, cooperative effects are taken into account. Then the distortion modes appearing in Eq. (3) are:

$$Q_1 = \frac{1}{\sqrt{3}}(Q_x + Q_y + Q_z) \quad \text{(A.1)}$$

$$Q_2 = \frac{1}{\sqrt{2}}(Q_x - Q_y) \quad \text{(A.2)}$$

$$Q_3 = \frac{1}{\sqrt{6}}(-Q_x - Q_y + 2Q_z) \quad \text{(A.3)}$$

And therefore, $Q_1 > 0$ represents an increase of the volume of the octahedra. Now, $\rho_i$ is the electronic density at site $i$. If one thinks intuitively about the electron-phonon interaction, either in terms of the Coulomb interaction between electrons and the negative charged oxygens, or in terms of hybridization between the Mn $d$ orbitals and the oxygen $p$ orbitals, an increase of the octahedra volume, or the Mn-O distances, should decrease the energy of the $d$ orbital.

In order to look at the differences of the two models (with or without the negative sign), we compare the (1,1) element of the 2×2 matrix describing the interaction in the $e_g$ subspace, with the two signs

$$Q_z\left(\frac{1}{\sqrt{3}} + \frac{2}{\sqrt{6}}\right) + (Q_x + Q_y)\left(\frac{1}{\sqrt{3}} - \frac{1}{\sqrt{6}}\right), \quad \text{(A.4)}$$

*versus,*

$$-(Q_x + Q_y)\left(\frac{1}{\sqrt{3}} + \frac{1}{\sqrt{6}}\right) - Q_z\left(\frac{1}{\sqrt{3}} - \frac{2}{\sqrt{6}}\right), \quad \text{(A.5)}$$

and similarly for the (2,2) element of the matrix. Off-diagonal terms are unaffected. We can see how the different signs change the role of a *planar breathing mode* ($Q_x + Q_y$) and the distortions in the $z$ direction $Q_z$. It also changes a global sign in both (1,1) and (2,2) elements that would have some consequences only if we are considering strained systems, and would predict strain with a wrong sign. This explains the small effect that the sign has on the results.

Only two more remarks: first, the coefficients are not exactly the same, therefore details, like the critical $\lambda$ for the appearance of an ordered phase, might vary slightly. And second, special care should be taken when imposing boundary conditions, that are different for different modes, such us freezing the apical oxygens in 2D calculations.

An approximation commonly made also minimizes the effect of the sign in front of $Q_1$. Mean field calculations for the undoped manganite ($x=0$, or 1 $e_g$ electron per site) show that the on-site Coulomb interaction at mean field level, can be modeled with an stiffer $Q_1$ elastic constant (usually taken to be 2 or larger).[30] With this approximation the expectation value of the $Q_1$ mode would be much less than the $Q_2$ and $Q_3$, especially at low temperatures.

In the present work, since we are interested in dopings much larger than $x=0$ (less $e_g$ electrons), the on-site Coulomb interaction is not expected to be as important, and therefore the same elastic constant is taken for all modes.

We have run simulations with the two signs and found similar results. We find that, as discussed in the article, the $Q_2$ is the most important mode, which is anyhow unaffected. In particular, choosing a positive sign in Eq. (3) and freezing the apical oxygens produces the same results as choosing a negative sign and imposing $\sum Q_z = 0$, for an almost identical range of $\lambda$. Since the former implies a smaller computational effort, it is the one used in the results of the body of the paper.